\documentclass[aps,prl,reprint,amsmath,amssymb]
{revtex4-1}
\usepackage{graphicx}
\begin{document}

\title{Enhanced magnetic ordering in Sm metal under extreme pressure}
\author{Y. Deng and J. S. Schilling}
\affiliation {Department of Physics, Washington University, St. Louis, Missouri 63130, USA}
\date{\today}
I
\begin{abstract}
The dependence of the magnetic ordering temperature $T_{\text{o}}$ of Sm
metal was determined through four-point electrical resistivity measurements
to pressures as high as 150 GPa. A strong increase in $T_{\text{o}}$ with
pressure is observed above 85 GPa. In this pressure range Sm ions alloyed in
dilute concentration with superconducting Y exhibit giant Kondo pair
breaking. Taken together, these results suggest that for pressures above $%
\sim $ 85 GPa Sm is in a highly correlated electron state, like a Kondo
lattice, with an unusually high value of $T_{\text{o}}.$ A detailed
comparison is made with similar results obtained earlier on Nd, Tb and Dy
and their dilute magnetic alloys with superconducting Y.
\end{abstract}

\maketitle

\section{\protect\bigskip Introduction}

Except for Ce, the local-moment magnetic state in elemental lanthanide
metals is highly stable. Under the application of sufficient
pressure, however, the magnetic state would be expected to destabilize. In
recent studies on the trivalent lanthanide metals Nd \cite{song}, Gd \cite%
{lim1}, Tb \cite{lim2}, and Dy \cite{lim1} the magnetic ordering
temperature $T_{\text{o}}$, with the exception of Gd, was found to rise
steeply to anomalously high values upon the application of extreme pressure.
In the same pressure range, alloying Nd, Tb, and Dy in dilute concentration
into superconducting Y resulted in a very large suppression of the superconducting 
transition temperature $T_{\text{c}}$, in the case of Y(Nd) the record value 
39 K/(at.\% Nd) \cite{song}. Such
high values are a signature of giant Kondo pair breaking, a sign that these
lanthanides may be approaching a magnetic instability. The anomalous rise in 
$T_{\text{o}}$ and the giant pair breaking thus appear to be related.

It is interesting to note that in the Kondo lattice model described by the
Doniach phase diagram \cite{doniach} $T_{\text{o}}$ is expected to first
increase with the magnitude of the negative covalent mixing exchange
coupling $J_{-}$ \cite{schrieffer} before passing through a maximum and
falling rapidly to the quantum critical point at 0 K (see Fig 9 in the
Discussion section). This occurs when the Ruderman-Kittel-Kasuya-Yosida
(RKKY) interaction \cite{kittel} is suppressed by Kondo spin screening.
Since the magnitude of $J_{-}$ normally increases under pressure \cite%
{schilling1}, in the Doniach picture $T_{\text{o}}$ versus pressure should
pass through a maximum and fall towards 0 K. This behavior was indeed recently
observed for elemental Nd metal by Song \textit{et al. }\cite{song}. In contrast, due to
the extreme stability of Gd's magnetic state with its half-filled 4$f^{7}$
configuration, even pressures to 1 or 2 Mbar are not sufficient to
bring Gd near a magnetic instability. Indeed, neither giant pair breaking in
Y(Gd) nor an anomalous rise in $T_{\text{o}}$ for Gd are observed at extreme
pressure \cite{lim1,fabbris}.

In view of the intriguing magnetic behavior in trivalent Nd metal and Y(Nd)
alloys at extreme pressure, an in-depth study of an additional light
trivalent lanthanide Sm, both as elemental metal and in the dilute magnetic
alloy Y(Sm), was undertaken. Sm metal crystallizes in the Sm-type ($\alpha $%
-Sm) structure at ambient pressure, transitioning to dhcp at 4 GPa, to fcc
at 14 GPa, to \textit{hR}24 (dfcc) at 19 GPa, to \textit{hP}3 at 37 GPa, and
finally to \textit{tI}2 at 91 GPa \cite{zhao,vohra}. These structural
transitions thus follow the regular trivalent lanthanide structure sequence
under pressure: hcp$\rightarrow $Sm-type$\rightarrow $dhcp$\rightarrow $fcc$%
\rightarrow $dfcc, a sequence generated by the increasing $d$ character in
the conduction band upon compression \cite{pettifor}.

Trivalent Sm assumes the configuration [Xe]4$f^{5}$ yielding for the free Sm$^{3+}$ ion 
in the ground state $^{6}$H$_{5/2}$,
with Land\'{e} $g$-factor $g_{J}$ = 2/7 and total angular momentum $J_{t}$ =
5/2. The effective magnetic moment of free Sm$^{3+}$calculated from Hund's
rules is $p_{eff}=$ 0.85 $\mu _{\text{B. }}$ However, magnetic susceptibility measurements on
paramagnetic Sm salts give $p_{eff}=$ 1.74 $\mu _{\text{B }}$\cite{blundell}. The difference
between the theoretical and experimental values is believed due to
contributions from low-lying excited states with different $J_{t}$ values.

In Sm metal the situation is more complicated since the crystalline
electric field and conduction-electron polarization significantly influence
the magnetic state of Sm$^{3+}$ \cite{adachi}. As a result of this
complexity, Sm metal exhibits a number of interesting physical phenomena.
Both the temperature-dependent heat capacity \cite{jennings} and electrical
resistivity \cite{alstad} have anomalies near 13 K and 106 K. The fact that
the temperature-dependent magnetic susceptibility of Sm has peaks near these
temperatures strongly suggests antiferromagnetic ordering \cite{mcewen}.
This was confirmed by Koehler and Moon from neutron diffraction experiments
on single crystalline $^{154}$Sm \cite{koehler}. They viewed the Sm-type structure 
with space group \textit{R}$\overline{3}$\textit{m}
as a combination of hexagonal and cubic sites where Sm$^{3+}$ ions at these
sites magnetically order at 106 K and 14 K, respectively.

In temperature-dependent resistivity measurements $R(T)$ a knee is observed
at the magnetic ordering temperature $T_{\text{o}}$ due to the loss of
spin-disorder scattering upon cooling. Dong \textit{et al.} \cite{dong}
measured resistivity on Sm to 43 GPa and found that the two ordering
temperatures move toward each other with increasing pressure, finally
merging together near 66 K at 8 GPa as Sm enters the dhcp phase. At higher
pressures $T_{\text{o}}$ increases rapidly to 135 K at 43 GPa. Johnson 
\textit{et al.} \cite{johnson} measured $R(T)$ on Sm to 47 GPa. They find
that the two ordering temperatures merge near 56 K at 10 GPa and increase
slowly up to the (\textit{hR}24$\rightarrow $\textit{hP}3) phase transition 
at 34 GPa where a second ordering temperature reportedly appears.

In this work four-point dc resistivity measurements are carried out on pure
Sm metal to $\sim $ 150 GPa using a diamond anvil cell. The two magnetic
ordering temperatures merge at 13 GPa after which $T_{\text{o}}(P)$
increases gradually to a maximum at 53 GPa, but then decreases and passes 
through a minimum at 85 GPa
followed by a sharp increase to $\sim $ 140 K at 150 GPa. Giant
superconducting pair breaking is also observed in dilute Y(Sm) alloys. Taken
together, both effects suggest that extreme pressure drives Sm to an
unconventional magnetic state, a state likely related to that observed
earlier in Nd, Tb, and Dy.

\section{\protect\bigskip Experimental Techniques}

Polycrystalline Sm samples for the high-pressure resistivity measurements
were cut from a Sm ingot. The dilute magnetic alloys of Y(Sm) were made by
argon arc-melting small amounts of Sm with Y (both Sm and Y 99.9 \% pure,
Ames Laboratory \cite{ames}). To enhance homogeneity the alloys were sealed
in glass ampules under vacuum and annealed at 600$^{\text{o}}$C for two
weeks. The concentrations of Sm for the four alloys as determined from x-ray
fluorescence analysis are: 0.15(2) at.\%, 0.40(3) at.\%, 0.83(4) at.\%, and
1.16(6) at.\%. Before arc-melting the nominal concentrations were 0.5 at.\%,
1.0 at.\%, 1.2 at.\%, and 2 at.\%, respectively. It follows that 30\% to
70\% of the Sm evaporated during arc-melting due to its relatively low
boiling point.

A diamond anvil cell (DAC) made of conventional and binary CuBe \cite{schilling}
was used to reach pressures to $\sim $150 GPa between two opposed diamond
anvils (1/6-carat, type Ia) with 0.18 mm diameter culets beveled at 7
degrees to 0.35 mm diameter. The force applied to the anvils was generated by a
stainless-steel diaphragm filled with He gas \cite{daniels}. The Re gasket (250 $\mu $m thick) was pre-indented to 30 
$\mu $m and a 90 $\mu $m diameter hole drilled through the center of the
pre-indentation area. A cBN-epoxy insulation layer was compressed onto the
surface of the gasket. Four Pt strips (4 $\mu $m thick) were then placed on
the insulation layer, acting as the electrical leads for the four-point
resistivity measurement. The Sm or Y(Sm) sample with dimensions 40 $\times $%
\ 40 $\times $ 4 $\mu $m$^{3}$ was then placed on the Pt strips. Further
details of the high-pressure resistivity techniques can be found
elsewhere \cite{lim1,shimizu}.

The DAC was inserted into an Oxford flow cryostat capable of varying
temperature from ambient to 1.3 K. Pressure was determined at room
temperature using the diamond vibron \cite{vibron}. Earlier resistivity
experiments by Song \textit{et al.} \cite{song} in an identical DAC using
both vibron and ruby manometers revealed an approximately linear pressure
increase of $\sim $ 30\% on cooling from 295 to 4 K. In the present experiments 
this calibration allows an estimate of the pressure at the magnetic or
superconducting transition temperatures from the vibron pressure at ambient
temperature.\bigskip

\section{Results of Experiment}

Four-point resistance measurements $R(T)$ were carried out on Sm in two runs
over the temperature and pressure ranges 1.3 - 295 K and 2 - 127 GPa
(measured at room temperature), respectively. The data from run 1 are shown
in Fig 1. For all pressures the resistance is seen to decrease upon cooling.
A kink or knee appears in $R(T)$ in the lower temperature range that results
from the progressive loss of spin-disorder scattering $R_{\text{sd}}(T)$ as
Sm orders magnetically. At 2 and 4 GPa two kinks are visible in the $R(T)$ curves; 
at higher pressures only one kink or knee appears. With increasing pressure the knee is seen to shift
in temperature and broaden; the broadening is due to the increasing pressure
gradient across the sample in the non-hydrostatic pressure environment. The
value of $T_{\text{o}}$ is determined from the intersection temperature of
two straight lines tracking $R(T)$ above and below the knee region, as
illustrated in Fig 1 for 27 GPa pressure. In most experiments the Sm sample was cooled
to $\sim 4$ K; however, considering both runs, at pressures 2, 9, 25, 51, 60, 86,
93, 97, and 127 GPa the sample was cooled to 1.3 K. In no experiment on Sm was
superconductivity, or even an onset to superconductivity, observed.

In Fig 2 the values of $T_{\text{o}}$ for Sm from runs 1 and 2 are plotted
versus pressure and compared to previous results from Dong \textit{et al.} 
\cite{dong} to 43 GPa and Johnson \textit{et al}. \cite{johnson} to 47 GPa.
In all experiments the two branches of $T_{\text{o}}$ are seen to merge near
13 GPa followed by an increase in $T_{\text{o}}$. In the present
experiments $T_{\text{o}}(P)$ passes through a maximum near\ 53
GPa, gradually decreasing to $\sim 60$ K near 85 GPa, before rising sharply
to $\sim 140$ K at 150 GPa. The report by Johnson \textit{et al.} \cite%
{johnson} that a second transition appears in the pressure range 35 - 50 GPa
could not be confirmed.

Due to the broadening of the resistivity knee under nonhydrostatic pressure,
the determination of the value of $T_{\text{o}}$ for Sm becomes
progressively more difficult in the upper pressure range. The same was true
for the other trivalent lanthanides Nd, Gd, Tb, and Dy studied previously \cite%
{song,lim1,lim2}. In particular, as here for Sm, a rapid upward shift of the
knee in $R(T)$ was also observed for Dy above 70 GPa pressure \cite{lim1}.
That the knee for Dy does indeed result from magnetic ordering over the
entire pressure range was recently confirmed by synchrotron M\"{o}ssbauer
spectroscopy (SMS) to 141 GPa \cite{bi}.

Independent information on the origin of the resistivity knee in Sm can be
gained by comparing the pressure dependence of the spin-disorder resistance $%
R_{\text{sd}}(P)$ for $T>T_{\text{o}}$ to that of $T_{\text{o}}(P)$ obtained
from the resistivity knee. As discussed in Ref \cite{lim1}, both $T_{\text{o}%
}$ \cite{taylor} and $R_{\text{sd}}$ \cite{daal} are proportional to $J^{2}N(E_{\text{F}})$, 
where $J$ is the exchange interaction between local
moment and conduction electrons and $N(E_{\text{F}})$ is the density of
states at the Fermi energy. A similarity between the pressure dependences $%
T_{\text{o}}(P)$ and $R_{\text{sd}}(P)$ is anticipated for the trivalent
lanthanide metals since their $spd$ conduction electron properties are
closely related. This similarity was indeed observed for Nd \cite{song}, Gd 
\cite{lim1}, Tb \cite{lim2}, and Dy \cite{lim1}; it would be interesting to
examine whether this also holds for Sm, together with Nd the second 
light lanthanide studied. From Fig 1 it is readily seen that
where the resistivity knee shifts under pressure to higher temperatures the
size of the resistivity drop-off below the knee also increases. A
semi-quantitative estimate of $R_{\text{sd}}$ is now attempted.

The total measured resistance is the sum of three terms, $R(T)=R_{\text{d}%
}+R_{\text{ph}}(T)+R_{\text{sd}}(T),$ where $R_{\text{d}}=R(0$ K$)$ is the
temperature-independent defect contribution. In the paramagnetic state in
the temperature region above the resistance knee, $R_{\text{sd}}(T)$ is
constant, taking on its maximum value $R_{\text{sd}}^{\text{max}}$, so that
the only temperature dependence comes from the phonon resistance $R_{\text{ph%
}}(T)$. To estimate $R_{\text{sd}}^{\text{max}},$ Colvin \textit{et al.} 
\cite{colvin} assumed that $R_{\text{ph}}(T)$\ depends linearly on
temperature, and extended a straight line fit to $R(T)$ for $T>T_{\text{o}}$
to 0 K with intercept $R_{\text{int}}$ and then subtracted off $R_{\text{d}}$
from this intercept. An example for this estimate is given in Fig 1 at 27
GPa where $R_{\text{sd}}^{\text{max}}=[R_{\text{int}}-R_{\text{d}}]=(115-18)$
m$\Omega =97$ m$\Omega $.

In Fig 3 $R_{\text{sd}}^{\text{max}}$ is plotted as a function of pressure.
Comparing Figs 2 and 3, a parallel behavior of the pressure dependences $T_{%
\text{o}}(P)$ and $R_{\text{sd}}^{\text{max}}(P)$ is indeed observed, thus
supporting the identification of the resistance knee with the onset of
magnetic ordering in Sm. Also included in Fig 3 is the quantity [$R$(290 K)
-- $R$(4 K)] that is seen to also qualitatively track $T_{\text{o}}$ versus
pressure. This suggests that the resistance from electron-phonon scattering
at room temperature does not change dramatically within the pressure range
of these experiments.

To examine whether the rapid rise in $T_{\text{o}}$ for pressures above 85
GPa might be related to an approaching instability in Sm's magnetic state,
Sm is alloyed in dilute concentration with Y, a high-pressure superconductor
having, compared to the trivalent lanthanides, closely similar conduction
electron properties and structural sequence under pressure \cite{wittig2}.
Under these circumstances the ability of the Sm ion to suppress Y's
superconductivity, the degree of pair breaking $\Delta T_{\text{c}}\equiv T_{%
\text{c}}$[Y] - $T_{\text{c}}$[Y(Sm)], can reveal valuable information about
the magnetic state of the Sm ion itself. This general observation was emphasized for
lanthanide ions by Maple \cite{maple}.

In the present experiment Y(Sm) alloys with differing dilute Sm
concentrations were studied at pressures to 180 GPa. Fig 4 shows the
superconducting transitions in four-point resistance measurements on Y(0.15
at.\% Sm) at selected pressures. As illustrated in this figure for the $R(T)$
data at 52 GPa, $T_{\text{c}}$ is defined as the temperature at which the
resistance transition reaches the halfway mark, whereas the intersection
point of two straight red lines defines $T_{\text{c}}^{\text{onset \ }}$, and $%
T_{\text{c}}^{\text{zero}}$ gives the temperature where the resistance
disappears. The fact that a typical total transition width is less than 2 K
gives evidence that the distribution of Sm ions in the alloys is
homogeneous. As seen from the data in Fig 4, $T_{\text{c}}$ increases
monotonically with pressure to 140 GPa, but then decreases to 180 GPa.

The dependence of $T_{\text{c}}$ on pressure for Y(Sm) alloys with Sm
concentrations 0.15, 0.40, 0.83, and 1.16 at.\% is shown in Fig 5. Below $%
\sim $ 40 GPa the $T_{\text{c}}(P)$ dependence for all four alloys tracks
that for pure Y. However, above $\sim $ 40 GPa a strong suppression sets in.
This suppression $\Delta T_{\text{c}}$ is so strong that for Y(1.16 at.\%
Sm) at pressures above 50 GPa $T_{\text{c}}$ lies below the temperature
range of this experiment (1.3 K). For the more dilute Y(0.15 at.\% Sm) and
Y(0.40 at.\% Sm) alloys, $T_{\text{c}}$ remains well above 1.3 K at all
pressures.

To allow a more meaningful comparison of the degree of superconducting pair
breaking $\Delta T_{\text{c}}$ for the different alloys, in Fig 6 $\Delta T_{%
\text{c}}$ is divided by the Sm concentration $c$ and then plotted versus
pressure for all alloys measured. Where they can be compared, the individual 
$\Delta T_{\text{c}}/c$ curves agree reasonably well and increase
monotonically with pressure, reaching the extremely high value of $\sim $ 40
K/at.\% Sm at 180 GPa, a value slightly higher than that found earlier for Y(0.4 at.\% Nd) 
\cite{song}. Both the giant pair breaking in Y(Sm) and the remarkable
increase of $T_{\text{o}}$ in Sm give evidence for unconventional physics in
Sm above 85 GPa.

\section{Discussion}

The present results on Sm and Y(Sm) alloys will now be
compared to those from earlier studies on the lanthanides Nd \cite{song}, Gd \cite{lim1}, 
Tb \cite{lim2}, and Dy \cite{lim1}. Going
from right to left across the lanthanide series (Lu to La) or by applying
pressure, one finds with few exceptions \cite{samudrala} the canonical rare-earth crystal
structure sequence hcp$\rightarrow $Sm-type$\rightarrow $dhcp$\rightarrow $%
fcc$\rightarrow $\textit{hR}24 believed to mainly arise from an increase in
the number $n_{\text{d}}$ of $d$-electrons in the conduction band \cite%
{pettifor}.

In the elemental lanthanide metals magnetic ordering arises from the indirect RKKY exchange
interaction between the magnetic ions. For a conventional lanthanide metal
with a stable magnetic moment, the magnetic ordering temperature $T_{\text{o}%
}$ is expected to scale with the de Gennes factor $(g-1)^{2}J_{t}(J_{t}+1)$,
modulated by the prefactor $J^{2}N(E_{\text{F}})$, where $J$ is the exchange
interaction between the 4$f$ ion and the conduction electrons, $N(E_{\text{F}%
})$ the density of states at the Fermi energy, $g$ the Land\'{e}-$g$ factor,
and $J_{t}$ the total angular momentum quantum number \cite{taylor}. 

In Fig 7(a) the dependence of the magnetic ordering temperature $T_{\text{o}%
} $ on pressure is shown for the four lanthanide metals Nd, Sm, Tb, and Dy.
Except for Nd,  $T_{\text{o}}(P)$ is seen to initially decrease rapidly with pressure, but then 
pass through a minimum and rise. $T_{\text{o}}(P)$ for Gd \cite{lim1} also shows this same initial behavior.
Since the de Gennes factor, in the absence of a magnetic instability or valence transition, 
is constant under pressure, the initial $T_{\text{o}}(P)$ dependence for the above lanthanides 
likely originates in the pressure dependence of the prefactor $J^{2}N(E_{\text{F}})$.
Electronic structure calculations for Dy support this conclusion \cite{jackson1,fleming}. 

The strong initial decrease in $T_{\text{o}}$ with
pressure in Sm (upper transition), Gd, Tb, and Dy occurs within the hcp and Sm-type phases. 
The minimum in $T_{\text{o}}(P)$ 
at approximately 20 GPa for Dy appears at somewhat lower pressures for Tb, Gd, and Sm, 
disappearing entirely for Nd. As discussed in some detail in Ref \cite{song}, this is consistent 
with an increase in the number of \textit{d} electrons in the conduction band going from Dy to Nd; 
the electronic structure and the crystal structures taken on by Nd resemble those of Dy but at 
a pressure approximately 30 - 40 GPa higher \cite{song}. The
systematic behavior for all five lanthanides Dy, Tb, Gd, Sm, and Nd in the region of pressure 
where the hcp, Sm-type, dhcp, and \textit{hR}24 structures occur, gives evidence that 
changes in the magnetic ordering temperature in this region
are mainly determined by corresponding changes in the properties
of the conduction electrons that mediate the RKKY interactions between the
magnetic lanthanide ions. 

It would seem helpful to propose that the $T_{\text{o}}(P)$ curves for each
element can be separated into two principal pressure regions: a
\textquotedblleft conventional\textquotedblright\ region at lower pressure
governed by the electronic properties of the conduction electrons and normal
positive exchange interactions $J_{+}$ between the lanthanide ion and the
conduction electrons, and an \textquotedblleft
unconventional\textquotedblright\ region at higher pressures where exotic
physics dominates leading to negative covalent-mixing exchange $J_{-}$ and
associated anomalous magnetic properties. In the \textquotedblleft
conventional\textquotedblright\ region the observed variations in $T_{\text{o%
}}(P)$ would be principally caused by changes in the prefactor $N(E_{\text{F}})J^{2}_{+}$
 with pressure. In the \textquotedblleft unconventional\textquotedblright\
pressure region highly correlated electron effects dominate leading to
anomalous magnetic properties, including anomalous $T_{\text{o}}(P)$ 
dependences and giant superconducting pair breaking in dilute magnetic alloys.

Although the properties of the conduction electrons and the magnetic state
of the lanthanide ion are intertwinned, the \textquotedblleft
conventional\textquotedblright\ and \textquotedblleft
unconventional\textquotedblright\ regions represent different physics, the
former being amenable through standard electronic structure
calculations, whereas the latter is only accessible through consideration of
strong highly correlated electron effects. The stability of the ion's
magnetic state is determined to a large extent by the exchange interactions 
\textit{within} a given lanthanide ion (Hund's rules). Once the
\textquotedblleft unconventional\textquotedblright\ rapid rise in $T_{\text{o%
}}$ with pressure sets in, it overpowers the \textquotedblleft
conventional\textquotedblright\ conduction electron behavior and determines $%
T_{\text{o}}(P)$. Since in Dy and Nd the \textquotedblleft
unconventional\textquotedblright\ region begins at a lower pressure, the
rapid rise in $T_{\text{o}}$ may prevent the \textquotedblleft
conventional\textquotedblright\ second minimum seen in Sm and Tb from
appearing in $T_{\text{o}}(P)$ for Dy or Nd.

A rough estimate of the boundary pressure where the \textquotedblleft
unconventional\textquotedblright\ $T_{\text{o}}(P)$ behavior may begin for a
given lanthanide is indicated by a vertical tick mark in Fig 7(a). In the
\textquotedblleft unconventional\textquotedblright\ region itself the $T_{%
\text{o}}(P)$ data curves have been given double thickness. There is a good
deal of arbitrariness for where this boundary is placed, particularly for Sm
and Tb where the second $T_{\text{o}}(P)$ minimum may well belong to the
\textquotedblleft unconventional\textquotedblright\ region, instead of the
\textquotedblleft conventional\textquotedblright\ region, as indicated by
the beginning of anomalous superconducting pair breaking in Y(Sm) or Y(Tb)
near the pressure for the second minimum in $T_{\text{o}}(P)$.

Focussing now on the anomalous rise in $T_{\text{o}}$ with pressure in
the \textquotedblleft unconventional\textquotedblright\ region in Fig 7(a),
we note that this rise is steepest for Nd but becomes progressively less
steep for Dy, Tb, and Sm. At least part of this reduction in steepness has
to do with the fact that the compressibility of the lanthanides decreases
significantly as pressure is increased. To bring out the physics more clearly, $%
T_{\text{o}}$ in Fig 7(b) is replotted versus the relative volume $V/V_{\text{o%
}}$. Different features in the respective curves are shifted to new relative
positions, but now it is seen that the sharp upturns in $T_{\text{o}}(P)$ have
nearly the same slope and are much steeper relative to the changes in the
\textquotedblleft conventional\textquotedblright\ region at lower pressures.
This points to a common mechanism for the upturn in these four lanthanides.

In Fig 8 the normalized pair breaking curve $\Delta T_{c}/c$\ for Y(Sm) from
Fig 6 is compared to those for the dilute magnetic alloys Y(Nd) \cite{song},
Y(Tb) \cite{lim2}, and Y(Dy) \cite{lim1}. For Y(Sm) and Y(Nd) the pair
breaking begins to increase rapidly at relatively low pressures compared to
Y(Tb) and especially Y(Dy). At least part of the reason for this is that the
Y host exerts lattice pressure on the light lanthanides Sm and Nd, but not
on Tb and Dy. This can be seen by comparing the respective molar volumes in
units of cm$^{3}$/mol: \ Y(19.88), Nd(20.58), Sm(19.98), Gd(19.90),
Tb(19.30), Dy(19.01) \cite{singman}. Without exception, the region of
pressure where $T_{\text{o}}(P)$ increases rapidly lies within the region of
pressure where the superconducting pair breaking $\Delta T_{c}/c$\ in the
corresponding dilute magnetic alloy with Y is anomalously large. Note also
that the maximum value of the slope of $\Delta T_{c}/c$ versus pressure in Fig 8 
is noticeably reduced for Y(Dy). At least part of this effect is due to the sizable
reduction in the compressibility of Y at higher pressures.

For the dilute magnetic alloy Y(Nd) the normalized pair breaking data in Fig
8 are seen to be reduced ($\Delta T_{c}/c$\ turns upwards) for pressures
above 160 GPa. Presumably the same effect\ would also be observed in Y(Sm),
Y(Tb), and Y(Dy) if the experiments were extended to even higher pressures.
This reduction in giant pair breaking seen in Y(Nd) at the highest pressures
was observed previously in dilute magnetic alloys La(Ce) \cite{wittig1},
La(Pr) \cite{wittig3}, and Y(Pr) \cite{fabbris,wittig4} and can be readily
accounted for in terms of Kondo pair-breaking theory \cite{zuckermann} where
the magnitude of the negative exchange interaction $J_{-}$ between the magnetic ions and the
conduction electrons increases with pressure. The appearance of such Kondo
physics in the dilute magnetic alloy suggests that the corresponding
concentrated system will likely show Kondo lattice, heavy Fermion, and
fluctuating valence behavior at higher pressures, eventually culminating in a
full increase in valence whereby one 4$f$ electron completely leaves its orbital and joins the
conduction band.

The well known Doniach model \cite{doniach} is often cited to account for
the dependence of the magnetic ordering temperature $T_{\text{o}}$ in a
Kondo lattice as a function of the magnitude of the negative exchange
parameter $J_{-}$ (see Fig 9). Whereas the upturn in $\Delta T_{c}/c$%
\ occurs above 160 GPa for Y(Nd), the downturn in $T_{\text{o}}(P)$ 
begins above 80 GPa (see Fig 7(a)) for Nd in its \textquotedblleft
unconventional\textquotedblright\ pressure region (double line width). The
rapid rise in $T_{\text{o}}(P)$ for Nd followed by its rapid downturn
resembles the dependence anticipated from the Doniach model \cite{song}. A similar $T_{%
\text{o}}(P)$ dependence would be expected for Sm, Tb and Dy if the
experiments were extended to even higher pressures.

The values of the pair-breaking parameter $\Delta T_{c}/c$\ for Nd and Sm
impurities in Y are surprisingly large - in fact, to our knowledge, the
largest ever reported. However, even more surprising is the sharp upturn in $%
T_{\text{o}}(P)$ where $T_{\text{o}}$ reaches values that appear to be much
higher than would have been possible had \textquotedblleft
unconventional\textquotedblright\ physics, such as Kondo physics, not been
operative. In the case of Dy, $T_{\text{o}}(P)$ extrapolates to values well
above room temperature, higher than any known value for an elemental
lanthanide metal at either ambient or high pressure \cite{lim1}. 

In summary, the magnetic properties of the light lanthanide Sm have been studied 
to extreme pressure and found to parallel those of another light lanthanide, Nd, 
as well as the heavy lanthanides Gd, Tb, and Dy. It appears that the magnetic phase 
diagram can be separated into two regions:  a low-pressure region where conventional 
changes in the electronic structure determine $T_{\text{o}}(P)$, and a high-pressure region 
where highly correlated electron effects dominate, leading to
such anomalous phenomena as unexpectedly high magnetic ordering temperatures and 
giant superconducting pair-breaking. The authors hope 
that this and previous work will lead to increased theoretical activity in this area.

\noindent \textbf{Acknowledgments.} The authors would like to thank A. K.
Gangopadhyay for his assistance in preparing the Y(Sm) alloys and R. A.
Couture for carrying out the x-ray fluorescence determination of the Sm
content in these alloys. Thanks are also due Daniel Haskel for his critical 
reading of the manuscript. This work was supported by the National
Science Foundation (NSF) through Grant No. DMR-1104742 and No. DMR-1505345
as well as by the Carnegie/DOE Alliance Center (CDAC) through NNSA/DOE Grant
No. DE-FC52-08NA28554.

\newpage

\begin{center}
{\LARGE Figure Captions}
\end{center}

\noindent Fig 1. Four-point resistance data $R(T)$ from run 1 for Sm metal
versus temperature on warming from 1.3 to 295 K at multiple pressures to 127
GPa (measured at room temperature). Knee in $R(T)$ at $T_{\text{o}}$ signals
onset of magnetic order (for example, at 27 GPa $T_{\text{o}}$ $\approx $ 61
K). Straight red line fitting data above knee for 27 GPa intercepts
resistance axis at 115 m$\Omega .$

\bigskip

\noindent Fig 2. Magnetic ordering temperature $T_{\text{o}}$ of Sm versus
pressure. Data from run 1 ($\bullet $), data from run 2 ($\blacktriangle $),
dotted line from Ref \cite{johnson}, dashed line from Ref \cite{dong}. Value
of pressure is estimated for temperature near $T_{\text{o}}$ (see text).
Question marks (?) accompany data points where evidence for magnetic
ordering is weak. Extended solid lines through data points are guides to the
eye. Crystal structures for Sm at top of graph determined to 189 GPa \cite%
{vohra}.

\bigskip

\noindent Fig 3. Plot of estimated maximum value of spin-disorder resistance 
$R_{\text{sd}}^{\text{max}}(P)$ versus pressure. $R_{\text{sd}}^{\text{max}}$
is estimated by subtracting defect resistance $R_{\text{d}}$ from intersection
point $R_{\text{int}}$ on resistance axis of straight-line fit to $R(T)$
data for $T>T_{\text{o}}$ (see text). Also shown is pressure dependence$\ $%
of [$R$(290 K) - $R$(4 K)] using data from Fig 1.

\bigskip

\noindent Fig 4. Four-point resistance of Y(0.15 at.\% Sm) alloy versus
temperature showing superconducting transition at various pressures to 180
GPa (estimated at low temperature). Intersection of two red straight lines
defines $T_{c}^{\text{onset}},$ midpoint of transition defines $T_{c},$
temperature where $R(T)\simeq $ 0 defines $T_{c}^{\text{zero}}.$

\bigskip

\noindent Fig 5. Superconducting transition temperature $T_{c}$ versus
pressure (estimated at low temperature) for Y and Y(Sm) alloys at four
different Sm concentrations. In all cases giant superconducting pair
breaking $\Delta T_{c}\equiv T_{c}[$Y$]-T_{c}[$Y(Sm)$]$ is observed. At top
of graph are crystal structures for superconducting host Y to 177 GPa \cite%
{samudrala}.

\bigskip

\noindent Fig 6. Superconducting pair breaking $\Delta T_{c}$ divided by
concentration $c$ of Sm in four Y(Sm) alloys plotted versus pressure. At top
of graph are crystal structures for superconducting host Y to 177 GPa \cite%
{samudrala}. Line through data is guide to the eye.

\bigskip

\noindent Fig 7. (a) Graph comparing lines through $T_{\text{o}}$ versus
pressure data for Nd, Sm, Tb, and Dy. Vertical tick marks separate regions
of \textquotedblleft conventional\textquotedblright\ (to left) from
\textquotedblleft unconventional\textquotedblright\ (to right) behavior in $%
T_{\text{o}}(P).$ (b) Data in (a) is replotted versus $V/V_{\text{o}},$
where $V_{\text{o}}$ is sample volume at ambient pressure, using measured
equations of state of Nd \cite{chesnut}, Sm \cite{zhao}, Tb \cite%
{cunningham}, and Dy \cite{patterson}. In both graphs lines with double
thickness mark regions where the magnetic ordering is \textquotedblleft
unconventional\textquotedblright .

\bigskip

\noindent Fig 8. Graph comparing relative superconducting pair breaking $\Delta T_{c}/c$ for dilute magnetic alloys
Y(Nd), Y(Sm), Y(Tb), and Y(Dy) versus pressure showing lines through data as
in Fig 6 for Y(Sm). At top of graph are crystal structures for
superconducting host Y to 177 GPa \cite{samudrala}.

\bigskip

\noindent Fig 9. \ Magnetic ordering temperature $T_{\text{o}}$\ vs absolute
value of negative exchange parameter $J$ according to the Doniach model \cite%
{doniach}. Since $T_{\text{o}}$ from the RKKY interaction increases as $%
J^{2},$ but is overtaken by the exponential increase of the Kondo
temperature $T_{\text{K}},$ the magnetic ordering is quenched.


\begin{thebibliography}{99}
\bibitem{song} J. Song, W. Bi, D. Haskel, and J. S. Schilling, Phys. Rev. B 
\textbf{95}, 205138 (2017).

\bibitem{lim1} J. Lim, G. Fabbris, D. Haskel, and J. S. Schilling, Phys.
Rev. B \textbf{91}, 045116 (2015).

\bibitem{lim2} J. Lim, G. Fabbris, D. Haskel, and J. S. Schilling, Phys.
Rev. B \textbf{91}, 174428 (2015).

\bibitem{doniach} S. Doniach, in \textit{Valence Instabilities and Related
Narrow-Band Phenomena}, edited by R. D. Parks (Plenum, New York, 1977), p.
169; S. Doniach, Physica B+C \textbf{91}, 231 (1977).

\bibitem{schrieffer} J. R. Schrieffer, J. Appl. Phys. \textbf{38}, 1143
(1967).

\bibitem{kittel} M. A. Ruderman and C. Kittel, Phys. Rev. \textbf{96}, 99
(1954).

\bibitem{schilling1} J. S. Schilling, Adv. Phys. \textbf{28}, 657 (1979).

\bibitem{fabbris} G. Fabbris, T. Matsuoka, J. Lim, J. R. L. Mardegan, K.
Shimizu, D. Haskel, and J. S. Schilling, Phys. Rev. B \textbf{88}, 245103
(2013).

\bibitem{zhao} Y. C. Zhao, F. Porsch, and W. B. Holzapfel, Phys. Rev. B 
\textbf{50}, 6603 (1994).

\bibitem{vohra} Y. Vohra, L. Akella, S. Weir, and G. A. Smith, Phys. Lett. A 
\textbf{158}, 89 (1991).

\bibitem{pettifor} J. C. Duthie and D. G. Pettifor, Phys. Rev. Lett. \textbf{%
38}, 564 (1977).

\bibitem{blundell} S. Blundell, in \textit{Magnetism in Condensed Matter}
(Oxford University Press, New York 2001) p. 34.

\bibitem{adachi} H. Adachi, H. Ino, and H. Miwa, Phys. Rev. B \textbf{56},
349 (1997).

\bibitem{jennings} L. D. Jennings, E. D. Hill, and F. H. Spedding, J. Chem.
Phys. \textbf{31}, 1240 (1959).

\bibitem{alstad} J. K. Alstad, R. V. Colvin, S. Legvold, and F. H. Spedding,
Phys. Rev. \textbf{121}, 1637 (1961).

\bibitem{mcewen} K. A. McEwen, P. F. Touborg, G. J. Cock and L. W. Roeland,
J. Phys. F: Metal Phys. \textbf{4}, 2264 (1974).

\bibitem{koehler} W. C. Koehler and R. M. Moon, Phys. Rev. Lett. \textbf{29}%
, 1468 (1972).

\bibitem{dong} W. Y. Dong, T. H. Lin, K. J. Dunn and C. N. J. Wagner, Phys.
Rev. B \textbf{35}, 966 (1987).

\bibitem{johnson} C. R. Johnson, G. M. Tsoi, and Y. K. Vohra, J. Phys.:
Condens. Matter \textbf{29}, 065801 (2017).

\bibitem{ames} Material Preparation Center, Ames Laboratory, U.S. Department
of Energy, Ames, Iowa, http://www.mpc.ameslab.gov.

\bibitem{schilling} James S. Schilling, in \textit{Proceedings of the 9th
AIRAPT International High Pressure Conf.}, Albany, New York, July 24-29,
1983, editors C. Homan, R.K. MacCrone and E. Whalley (North-Holland, N.Y.,
1984); Mat. Res. Soc. Symp. Proc. \textbf{22}, 79 (1984).

\bibitem{daniels} W. B. Daniels and W. Ryschkewitsch, Rev. Sci. Instr. 
\textbf{54}, 115 (1983).

\bibitem{shimizu} K. Shimizu, K. Amaya, and N. Suzuki, J. Phys. Soc. Jpn. 
\textbf{74}, 1345 (2005).

\bibitem{vibron} Y. Akahama and H. Kawamura, J. Appl. Phys. \textbf{100},
043516 (2006).

\bibitem{bi} W. Bi, J. Song, Y. Deng, P. Materne, J. Zhao, E. E. Alp, M. Y.
Hu, D. Haskel, Y. Lee, and J. S. Schilling (unpublished).

\bibitem{taylor} See, for example, K. N. R. Taylor and M. I. Darby, \textit{%
Physics of Rare Earth Solids} (Chapman and Hall, London, 1972).

\bibitem{daal} H. J. van Daal and K. H. J. Buschow, Solid State Commun. 
\textbf{7}, 217 (1969).

\bibitem{colvin} R. V. Colvin, S. Legvold, and F. H. Spedding, Phys. Rev. 
\textbf{120}, 741 (1960).

\bibitem{wittig2} J. Wittig, Phys. Rev. Lett. \textbf{24}, 812 (1970); J. J.
Hamlin, V. G. Tissen, and J. S. Schilling, Physica C (Amsterdam) \textbf{451}%
, 82 (2007).

\bibitem{maple} M. B. Maple, Appl. Phys. \textbf{9}, 179 (1976).

\bibitem{samudrala} G. K. Samudrala, G. M. Tsoi, and Y. K. Vohra, J. Phys.
Condens. Matter \textbf{24}, 362201 (2012).

\bibitem{jackson1} D. D. Jackson, V. Malba, S. T. Weir, P. A. Baker, and Y.
K. Vohra, Phys. Rev. B \textbf{71}, 184416 (2005).

\bibitem{fleming} G. S. Fleming and S. H. Liu, Phys. Rev. B \textbf{2}, 164
(1970); S. H. Liu, Phys. Rev. \textbf{127}, 1889 (1962).

\bibitem{chesnut} G. N. Chesnut and Y. K. Vohra, Phys. Rev. B \textbf{61},
R3768 (2000).

\bibitem{cunningham} N. C. Cunningham, W. Qiu, K. M. Hope, H.-P. Liermann,
and Y. K. Vohra, Phys. Rev. B \textbf{76}, 212101 (2007).

\bibitem{patterson} R. Patterson, C. K. Saw, and J. Akella, J. Appl. Phys. 
\textbf{95}, 5443 (2004).

\bibitem{singman} C. N. Singman, J. Chem. Ed. \textbf{61}, 137 (1984).

\bibitem{wittig1} M. Maple, J. Wittig, and K. Kim, Phys. Rev. Lett. \textbf{%
23}, 1375 (1969).

\bibitem{wittig3} J. Wittig, Phys. Rev. Lett. \textbf{46}, 1431 (1981).

\bibitem{wittig4} J. Wittig, \textit{Valence Instabilities}, edited by P.
Wachter and H. Boppart (North-Holland, Amsterdam, 1982), p. 427.

\bibitem{zuckermann} M. Zuckermann, Phys. Rev. \textbf{168}, 390 (1968); E. M%
\"{u}ller-Hartmann and J. Zittartz, Z. Physik \textbf{234}, 58 (1970).

\end{thebibliography}
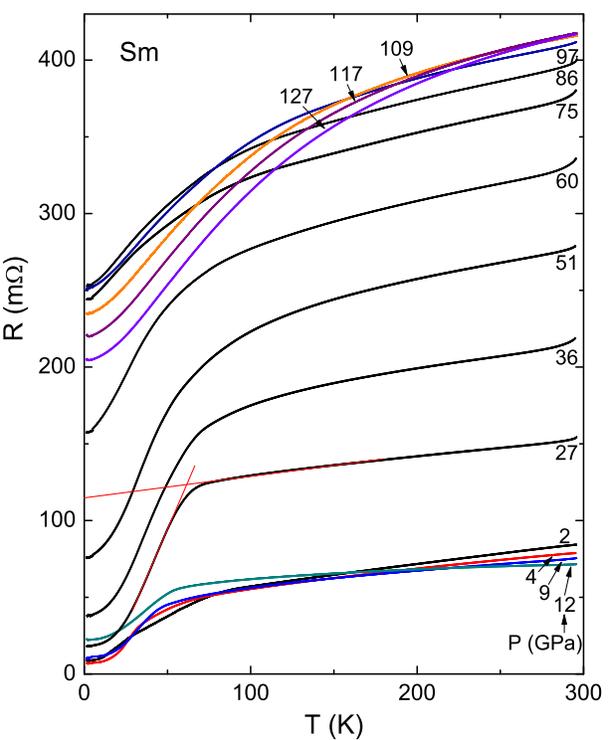
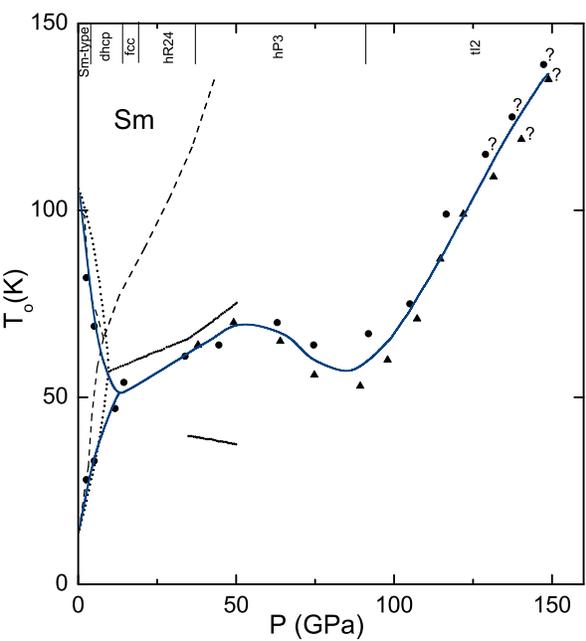
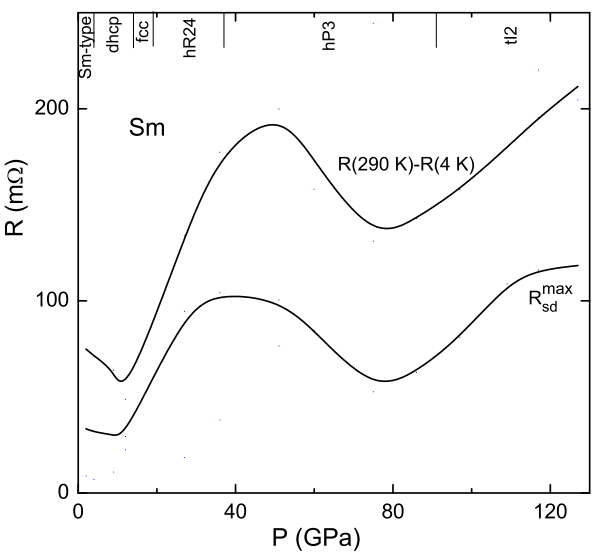
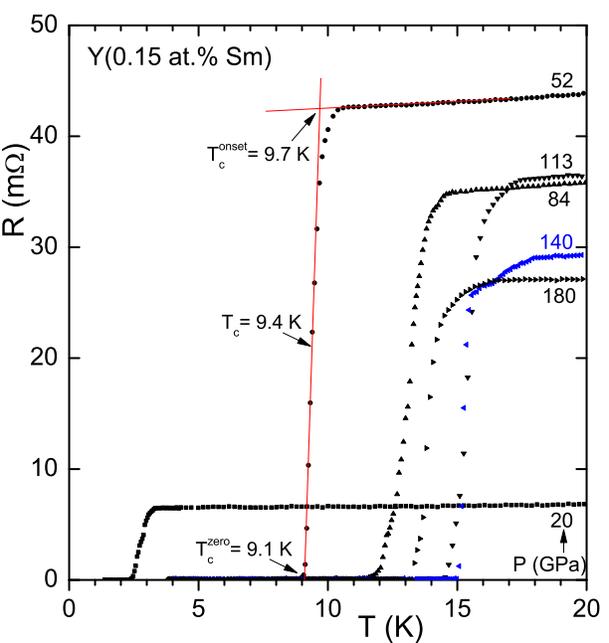
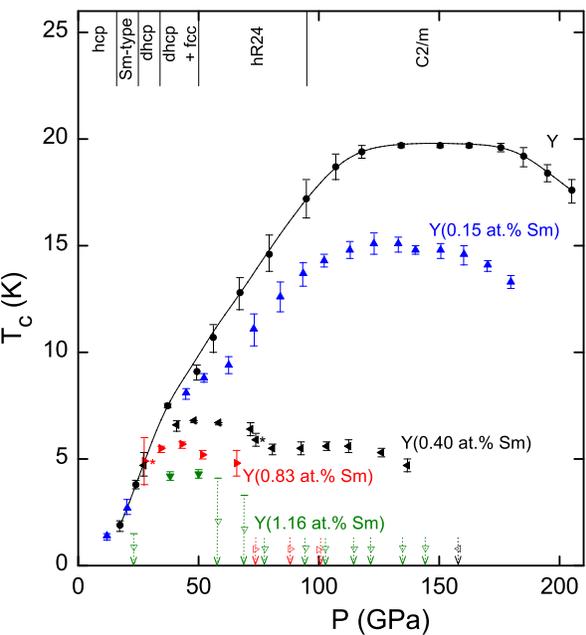
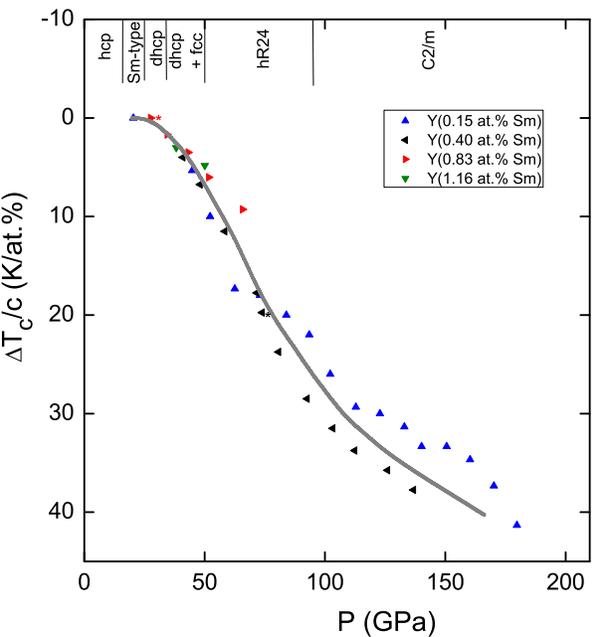
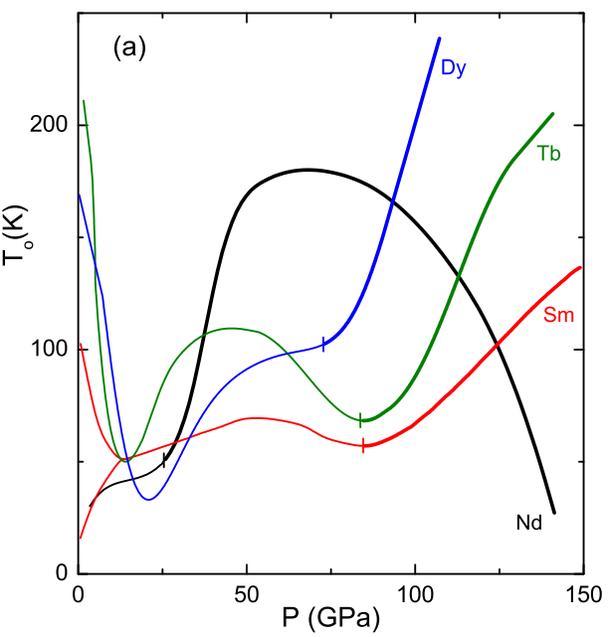
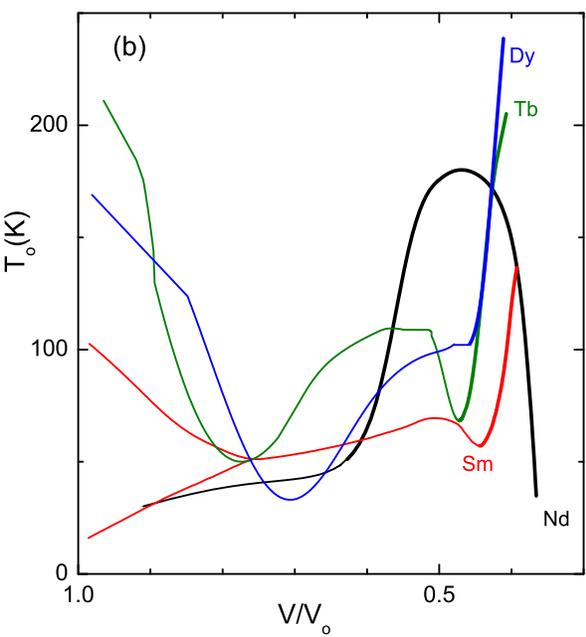
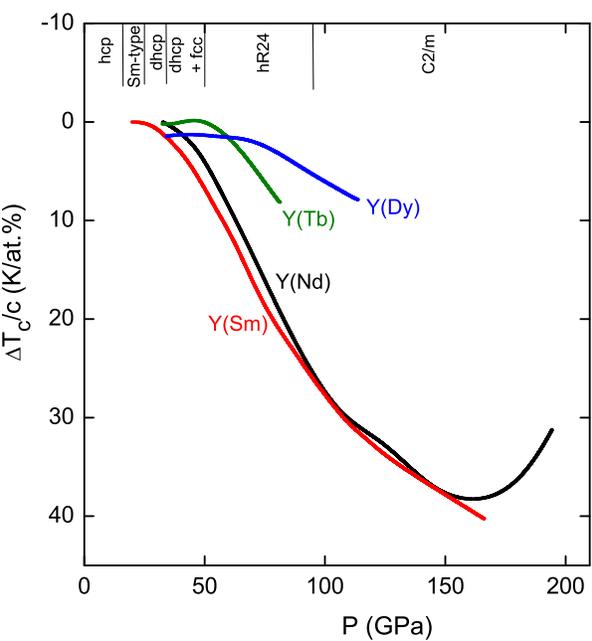
\end{document}